%% file: main.tex
\documentclass[10pt,twocolumn,letterpaper]{article}

\usepackage{iccv}              % To produce the CAMERA-READY version

\usepackage{amsthm, amssymb, amsmath}

\usepackage{color}
\usepackage{xcolor}
\usepackage{tcolorbox}
\usepackage{comment}
\usepackage{array}
\usepackage{multirow}
\usepackage{tabularx}
\usepackage{xcolor}
\usepackage{pgfplots}
\pgfplotsset{compat=1.18}
\usepackage{caption}
\usepackage{subcaption}

\usepackage[linesnumbered,ruled]{algorithm2e}

\usepackage{times}
\usepackage{latexsym}
\usepackage{pgfplots}
\usepackage{pgf-pie}
\usetikzlibrary{arrows.meta, patterns}
\usepackage[T1]{fontenc}
\usepackage[utf8]{inputenc}
\usepackage{tikz}
\usetikzlibrary{positioning}
\usepackage{microtype}
\usepackage{pgf-pie}
\usepackage{inconsolata}
\usepackage{graphicx}
\usetikzlibrary{decorations.pathreplacing}
\usepackage[justification=centering]{caption}
\usepackage{comment}
\usepackage{booktabs}
\theoremstyle{definition}

\newtheorem{example}{Example}

\definecolor{iccvblue}{rgb}{0.21,0.49,0.74}
\usepackage[pagebackref,breaklinks,colorlinks,allcolors=iccvblue]{hyperref}

\def\paperID{*****} % *** Enter the Paper ID here
\def\confName{ICCV}
\def\confYear{2025}

\title{Trustworthy Medical Imaging with Large Language Models: \\A Study of Hallucinations Across Modalities}

\author{Anindya Bijoy Das\\
The University of Akron\\
Akron, OH, USA\\
{\tt\small adas@uakron.edu}
\and
Shahnewaz Karim Sakib\\
University of Tennessee at Chattanooga\\
Chattanooga, TN, USA\\
{\tt\small ssakib1@utc.edu}
\and
Shibbir Ahmed\\
Texas State University\\
San Marcos, TX, USA\\
{\tt\small shibbir@txstate.edu}
}

\begin{document}
\maketitle
\input{sec/0_abstract}    
\vspace{-0.1 in}
\input{sec/1_intro}

\input{sec/2_literature}
\input{sec/3_finalcopy}
\input{sec/4_results}
\vspace{-0.1 in}
\input{sec/5_conclusion}
{
    \small
    \bibliographystyle{ieeenat_fullname}
    \bibliography{main}
}

\end{document}

%% file: sec/0_abstract.tex
\begin{abstract}
Large Language Models (LLMs) are increasingly applied to medical imaging tasks, including image interpretation and synthetic image generation. However, these models often produce hallucinations, which are confident but incorrect outputs that can mislead clinical decisions. This study examines hallucinations in two directions: image to text, where LLMs generate reports from X-ray, CT, or MRI scans, and text to image, where models create medical images from clinical prompts. We analyze errors such as factual inconsistencies and anatomical inaccuracies, evaluating outputs using expert informed criteria across imaging modalities. Our findings reveal common patterns of hallucination in both interpretive and generative tasks, with implications for clinical reliability. We also discuss factors contributing to these failures, including model architecture and training data. By systematically studying both image understanding and generation, this work provides insights into improving the safety and trustworthiness of LLM driven medical imaging systems.
\end{abstract}
\vspace{-0.1 in}

%% file: sec/1_intro.tex
\vspace{-0.1 cm}
\section{Introduction}
\label{sec:intro}

Large Language Models (LLMs) \cite{fink2023potential} have shown remarkable capabilities across a range of natural language processing tasks like summarization, question answering, translation, and content generation, enabling advances in education, law, and research. 
In the medical domain, LLMs hold significant promise \cite{ahmed2025medical} for improving clinical workflows by automating documentation, assisting in decision support, and enabling patient-specific summarization of complex medical information \cite{pal2023med}. 
When integrated with imaging data, LLMs can aid in generating radiology reports, explaining imaging findings, and synthesizing realistic images for education and training. These applications could enhance diagnostic efficiency, reduce clinician burden \cite{asgari2025framework}, and improve communication with patients. However, realizing these benefits requires careful evaluation of the accuracy, reliability, and safety of LLM outputs, especially in high-stakes healthcare settings.

Consider the case of brain MRI: essential for evaluating conditions like stroke, tumors, edema, and demyelination. A key finding here is midline shift \cite{li2025towards}, indicating displacement of brain structures due to mass effect from hemorrhage, tumor, or swelling.
%Consider the case of brain MRI, a critical imaging technique widely used to evaluate neurological conditions such as stroke, tumors, edema, and demyelinating diseases. One key finding that radiologists assess is midline shift—a displacement of brain structures caused by mass effect from hemorrhage, tumor, or swelling. 
The degree of midline shift is clinically significant, as even a few millimeters can indicate elevated intracranial pressure or impending herniation. 
If an LLM used for image interpretation fails to detect or misestimates midline shift, it may produce a report that downplays a life-threatening condition, potentially leading to harmful clinical decisions.
%If an LLM used for image interpretation fails to detect or incorrectly estimates the extent of midline shift, it may produce a report that downplays a life-threatening condition, potentially leading to harmful clinical decisions. 
On the generative side, if an LLM is prompted to produce an MRI scan showing a 6 mm midline shift but generates an inaccurate or subtle image, it may hinder medical education or misguide diagnostic model training. 

 \begin{figure}[t]
  \centering
  \captionsetup{justification = justified, singlelinecheck = false}
   \includegraphics[width=0.93\linewidth]{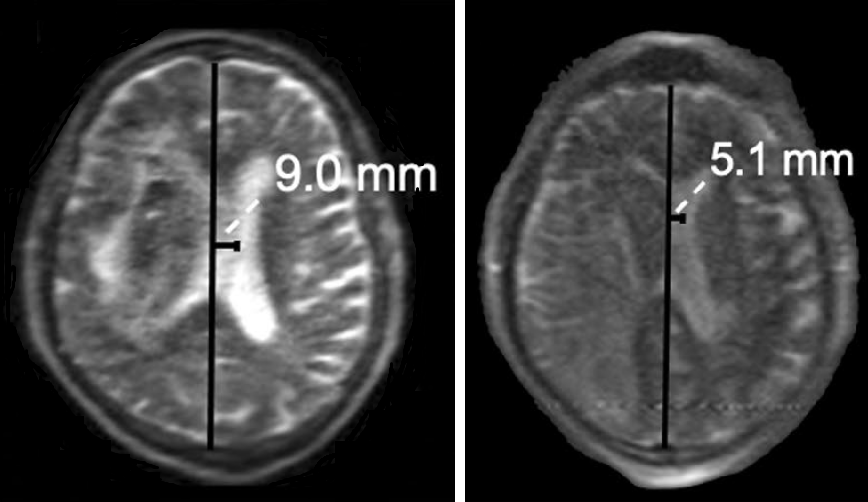}
      \label{fig:onecol}
      \vspace{-0.1 in}
   \caption{\small Example of midline shift measurements on portable MRI (pMRI) for a 81-year-old male having right intracerebral hemorrhage with a midline shift  9.0 mm (left) and a 71-year-old male having right M1 occlusion with a midline shift 5.1 mm (right).}
   \vspace{-0.19 in}
\end{figure}

Despite their potential, LLMs are prone to generating hallucinations, i.e., outputs that are fluent and confident but factually incorrect or unsupported by input data. 
In medicine, such errors can misrepresent patient data, introduce inaccuracies, or fabricate plausible-sounding diagnoses \cite{das2025hallucinations}.
%In medical applications, these hallucinations can be particularly harmful, as they may introduce inaccurate clinical information, misrepresent patient data, or fabricate plausible sounding diagnoses. 
Hallucinations can arise due to limitations in training data, overgeneralization, or the model's tendency to complete patterns based on learned linguistic structures rather than grounded evidence \cite{hardy2024rextrust}. 
In image to text tasks, this may lead to erroneous interpretations of medical images, while in text to image generation, the model might synthesize anatomically implausible or clinically misleading content. These errors, if unrecognized, risk spreading misinformation and jeopardizing patient safety \cite{yan2024med}.
%Therefore, understanding how and why hallucinations occur is critical for deploying LLMs in sensitive medical contexts and for developing safeguards that promote trustworthy outputs.

% In this work, we systematically investigate hallucinations in large language model applications to medical imaging, focusing on both image to text and text to image tasks. First, we analyze how LLMs generate clinical descriptions from medical images, using datasets across multiple modalities including chest X-rays, CT scans, and brain MRIs. We identify common hallucination patterns and assess the accuracy of generated reports against ground truth annotations and expert assessments. Next, we explore the reverse direction—generating medical images from textual prompts—evaluating whether the outputs faithfully represent the specified clinical findings and anatomical details. By examining hallucinations across both tasks, our study provides a comprehensive understanding of the limitations and risks involved in LLM-driven medical imaging pipelines. This work offers critical insights for the research community and healthcare practitioners by highlighting failure modes, suggesting evaluation strategies, and paving the way toward more trustworthy and clinically robust applications of generative AI in medicine.

In this work, we systematically investigate hallucinations in large language model (LLM) applications to medical imaging, focusing on both image-to-text and text-to-image tasks. Our key contributions are summarized below:

\begin{itemize}
    \item We analyze LLM-generated clinical descriptions from medical images across modalities such as chest X-rays, CT scans, and brain MRIs, identifying common hallucination patterns and evaluating report accuracy against ground truth and expert assessments.
    \item We evaluate text-to-image generation to determine whether the synthesized medical images faithfully represent the intended clinical findings, anatomical details, and align with plausible real-world clinical outcomes.
    \item We conduct numerical experiments across image interpretation, image classification, and image generation tasks to systematically identify and analyze hallucinations in LLM-based medical imaging workflows.
\end{itemize}
Together, these contributions can offer a comprehensive foundation for understanding and reducing hallucinations in LLM-driven medical imaging systems.

\section{Motivating Scenarios}
\vspace{-0.05 in}

To motivate our study, this section presents two examples  exposing critical limitations of current multimodal LLMs in medical imaging. The first one highlights issues in image interpretation, while the second reveals failures in image generation. These cases underscore the need for deeper analysis and robust safeguards in trustworthy medical AI.

\vspace{-0.05in}
\begin{example}
Consider the case of chest X-rays, a common imaging modality used to assess lung and pleural conditions. One important finding radiologists look for is pleural effusion \cite{li2024prompt}, i.e., the abnormal accumulation of fluid in the pleural space surrounding the lungs. Pleural effusion can occur on the left, right, or both sides (bilateral), and its severity is typically categorized as absent, mild, moderate, or large. Accurate detection of this condition is essential for diagnosing underlying causes such as heart failure, infection, or malignancy. If an LLM tasked with interpreting chest X-rays fails to identify the presence or severity of pleural effusion, it may generate a misleading report that could result in delayed or incorrect treatment. Similarly, if an LLM is prompted to create a chest X-ray image showing moderate right-sided pleural effusion but produces an anatomically inconsistent or incorrect image, it can misinform both clinical training and algorithmic development. 

\begin{figure}[t]
  \centering
    \captionsetup{justification = justified, singlelinecheck = false}
   \includegraphics[width=0.99\linewidth]{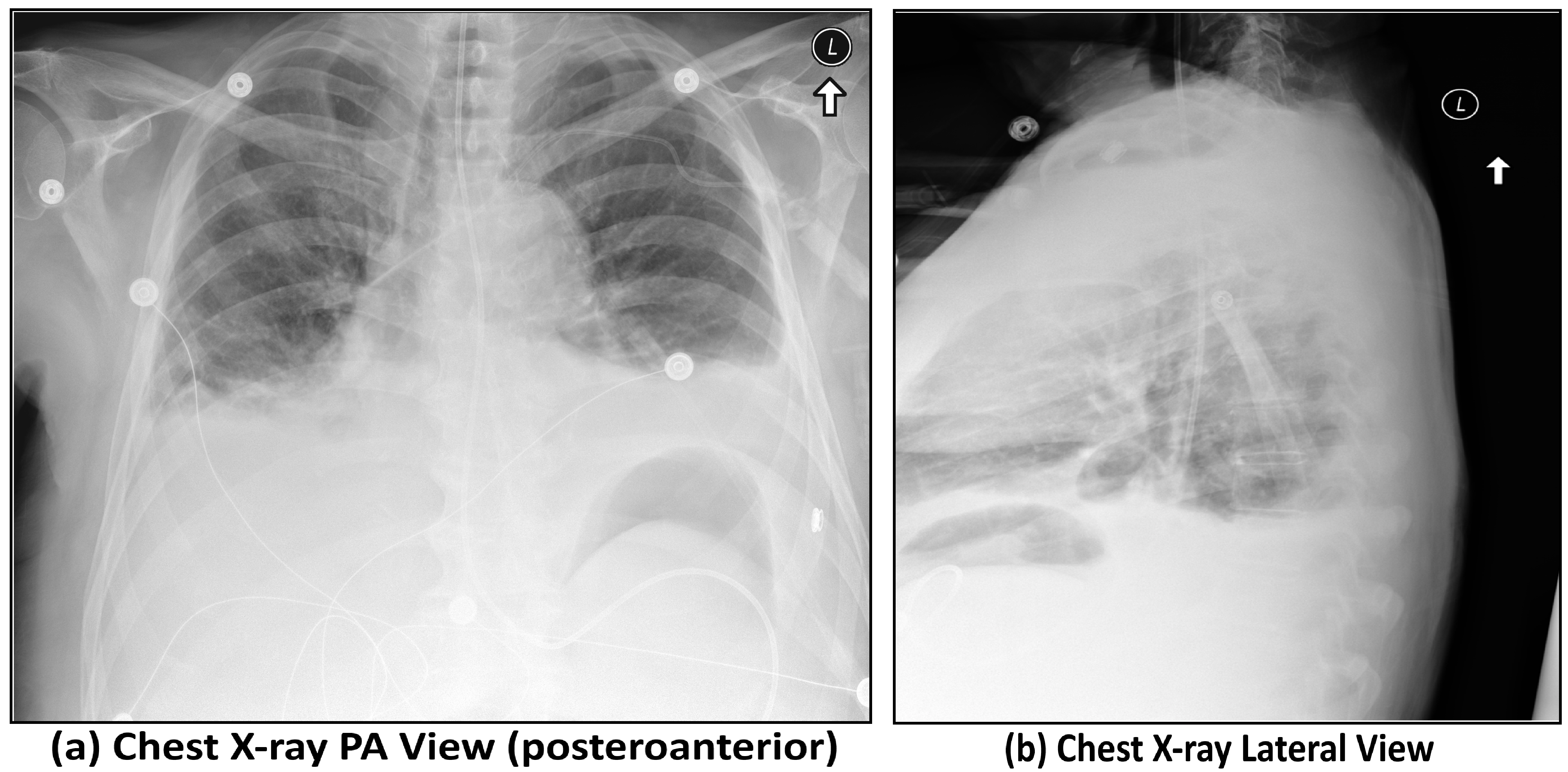}
   \vspace{-0.1 in}
   \caption{\small Chest X-ray (a) PA view and (b) lateral view with bilateral moderate to large pleural effusions.}
   \label{fig:onecol}
\end{figure}

\begin{table}
  \centering
    \caption{\small LLMs' performance to detect Pleural Effusion}
\vspace{-0.1 in}
  \begin{tabular}{@{}lc@{}}
    \toprule
    LLM & Decision  \\
    \midrule
    Grok 3 & right-sided, large. \\
    GPT 4o & bilateral, moderate to large \\
    Gemini 2.5 Flash & bilateral, moderate to large \\
    Claude Sonnet 4 & No \\
    Copilot & left-sided, large \\
    Perplexity & left-sided, large  \\
    \bottomrule
  \end{tabular}
\vspace{-0.1in}
  \label{tab:example}
\end{table}    
\end{example}

\vspace{-0.05in}
\begin{example}
In this example, we prompted various multimodal LLMs to generate a chest X-ray that clearly shows toe fractures. This request is intentionally inconsistent, as toe fractures cannot appear in chest X-rays. GPT-4o \cite{achiam2023gpt} nonetheless proceeds to generate an image—erroneously displaying fractured finger bones, likely from a hand, while still labeling it as a chest X-ray. This response not only reflects a fundamental misunderstanding of medical imaging anatomy but also reveals a concerning hallucination that may mislead users.  In contrast, Gemini-2.5 Flash \cite{team2023gemini} handles the request more responsibly: it produces two distinct images and provides clarification that a chest X-ray cannot capture toe fractures. This discrepancy highlights the variability in model behavior and emphasizes the need for rigorous evaluation of hallucinations in medical image generation.

\begin{figure}[t]
  \centering
    \captionsetup{justification = justified, singlelinecheck = false}
   \includegraphics[width=0.95\linewidth]{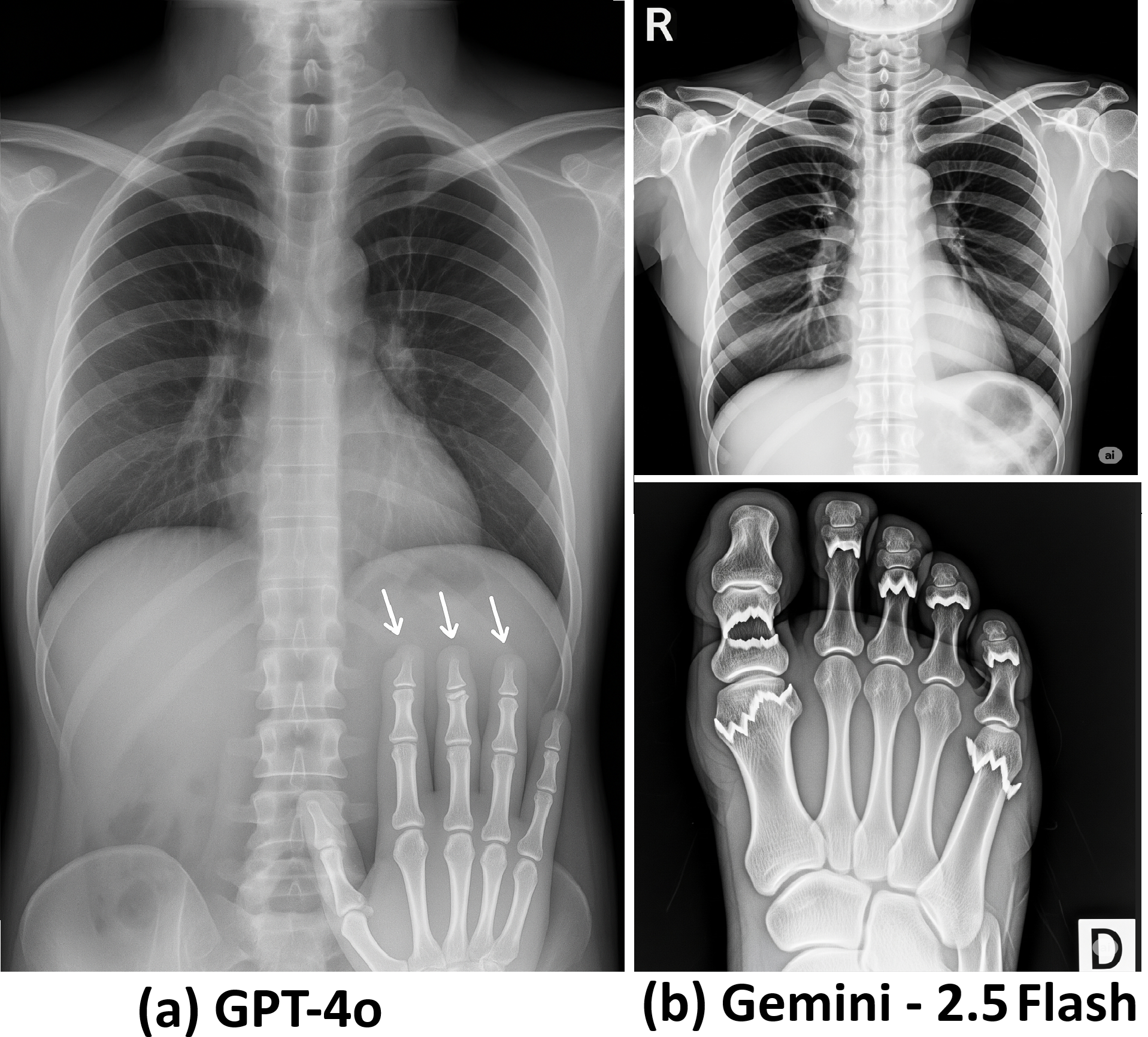}
   \vspace{-0.1 in}
   \caption{\small Model responses to the prompt: ``Generate a chest X-ray that clearly shows toe fractures.'' (a) GPT-4o hallucinates, overlaying fingers (not even toes) on a chest X-ray. (b) Gemini 2.5 Flash returns separate, anatomically correct chest and foot X-rays.}
   \label{fig:onecol}
\end{figure}

\end{example}

These risks underscore the importance of systematically identifying and mitigating hallucinations in both interpretive and generative applications of LLMs in trustworthy medical imaging \cite{ahmad2023creating}. Such hallucinations can lead to clinically misleading outputs, undermine user trust, and pose potential safety concerns in high-stakes environments. To contextualize these challenges, we examine existing literature that specifically addresses hallucination phenomena in large language and vision-language models, with an emphasis on their manifestations in medical imaging tasks.

%% file: sec/2_literature.tex
\section{Literature Survey}

%We now review relevant literature to contextualize the challenges highlighted earlier, focusing on hallucination detection, multimodal LLMs, and medical image analysis. Prior work on LLM hallucinations in medical imaging can be grouped into two main areas: (1) image interpretation and (2) image generation. We review recent advances in both domains below.
We now review relevant literature to contextualize the challenges highlighted earlier, focusing on hallucination detection, multimodal LLMs, and medical image analysis. Prior work on LLM hallucinations in medical imaging falls into two main areas: (1) image interpretation and (2) image generation. We review recent advances in both domains below.

\subsection{Hallucinations in Image Interpretation}

% Recent research highlights the prevalence of hallucinations in large language models (LLMs) and large vision-language models (LVLMs) when interpreting medical images. Gu et al. introduced MedVH \cite{gu2024medvh}, a systematic evaluation framework for hallucinations in LVLMs, emphasizing the risk of unverified or misleading outputs in clinical decision-making. Similarly, Chu et al. demonstrated that even advanced multimodal LLMs are prone to hallucination, particularly when handling complex vision-and-text tasks in healthcare, and proposed Visual Retrieval-Augmented Generation (V-RAG) as a mitigation strategy \cite{chu2025reducing}. The work by Kim et al. further corroborated these findings, showing that hallucinations can manifest as spurious structures in image interpretation, with diffusion-based frameworks offering partial remedies \cite{kim2025tackling}.

Recent research works highlight the prevalence of hallucinations in LLMs and large vision-language models (LVLMs) when interpreting medical images. The authors in \cite{gu2024medvh} introduced a systematic evaluation framework for hallucinations in LVLMs, emphasizing the risk of unverified or misleading outputs in clinical decision-making. Similarly, the work in \cite{yan2024med} demonstrated that even advanced multimodal LLMs are prone to hallucination, particularly when handling complex vision-and-text tasks in healthcare. The study in \cite{kim2025tackling} further corroborated these findings, showing that hallucinations can manifest as spurious structures in image interpretation, with diffusion-based frameworks offering partial remedies. The work in \cite{zhang2024radflag} highlights similar hallucination issues across multiple VLMs.

The clinical implications of hallucinations are significant, as misleading outputs can compromise patient safety and erode trust in AI systems.  Repeated errors may reduce reliance on these technologies by both clinicians and patients. To address this, strategies like prompt engineering have been proposed to guide AI-generated analyses, such as radiology reports, and reduce major errors \cite{kim2025medical}. Beyond general hallucinations, measurement hallucinations present distinct challenges in quantitative imaging. The study in \cite{heiman2025factchexcker} addressed this by integrating vision models for accurate measurements in chest X-ray interpretation, improving clinical reliability. Fact-checking frameworks like those in \cite{hardy2024rextrust} and \cite{heiman2025factchexcker} have also been adapted to medical outputs, using white-box detection and hidden state analysis to identify ungrounded claims and better assess hallucination risk.

There have also been growing efforts tackle these issues in AI systems through a combination of manual and automated approaches. The study in \cite{asgari2025framework} developed a clinician-led evaluation framework to label hallucinated and omitted content in LLM-generated clinical notes, highlighting ongoing challenges in ensuring factual accuracy. These findings align with broader observations that hallucinations are inherent to LLMs and influenced by training data, model design, and prompting \cite{gu2024medvh}. Complementing manual assessments, the work in \cite{chu2025reducing} introduced V-RAG, a retrieval-augmented generation approach that integrates text and visual data from retrieved images to improve entity grounding. Combined with entity probing and multimodal fine-tuning, this method enhances factual consistency and reduces clinically significant hallucinations, as further supported by findings in \cite{hardy2024rextrust}.

\subsection{Hallucinations in Image Generation}
Image generation differs from image interpretation, as it synthesizes visual content from prompts rather than analyzing existing images. However, generative models are broadly susceptible to hallucinations \cite{lim2024addressing}, often producing content that deviates from the intended prompt or includes unrealistic elements \cite{das2025breaking}. In medical settings, such hallucinations may lead to anatomically implausible or clinically misleading visuals. This poses serious concerns for accurate diagnostic training, effective medical education, and the development of reliable downstream AI tools.

%Similarly, hallucinations also significantly affect medical image generation, where LLM-driven models synthesize medical images from textual prompts. 
Recent studies have highlighted frequent occurrences of anatomically incorrect or clinically misleading visual outputs. For example, the work in \cite{zuo2024medhallbench} introduced benchmarks to systematically evaluate hallucinations in synthetic medical images, demonstrating that even specialized models can fabricate unrealistic structures or textures. Similarly, the study in \cite{tivnan2024hallucination} proposed a hallucination index for diffusion-based reconstruction in medical imaging, highlighting challenges with fidelity and artifact generation.

Several research efforts have focused on mitigating hallucinations in medical image generation. Among these, diffusion-based methods \cite{kiritani2024mitigating} have been proposed to identify and selectively refine problematic image regions, thereby reducing anatomical inaccuracies and improving overall clinical reliability.
The work in \cite{hsieh2025dall} addresses hallucinations by employing a multistep process that synthesizes data from medical images and text reports, ensuring generated visuals closely align with clinically relevant characteristics and thereby reducing misleading outputs.
Additionally, RAG can mitigate hallucinations by grounding outputs in retrieved medical knowledge, enhancing consistency and control, which is crucial for reliable clinical image generation \cite{rodriguez2024leveraging}.
Collectively, these works underscore ongoing challenges and emphasize the need for improved evaluation metrics, targeted refinement strategies, and careful model tuning to achieve clinically trustworthy image synthesis.

%% file: sec/3_finalcopy.tex
\section{Hallucination Analysis in Medical Imaging}
To better understand how large language models (LLMs) behave in medical imaging tasks, in this section, we conduct a detailed analysis of hallucinations across both image interpretation and image generation. 

\subsection{Image Interpretation}
First, we evaluate how LLMs interpret and reason about medical images. We begin with a classification task, where models are asked to identify general diagnostic categories (e.g., presence or absence of pathology) from medical images. Then, we investigate whether LLMs can detect and describe specific clinical events or findings within an image, such as fractures, effusions, or abnormal growths. These evaluations reveal the extent to which LLMs understand medical visual content and where hallucinations tend to arise.

\subsubsection{Medical Image Classification}
\label{sec:zeroshot}
{\bf Zero-shot setting:} We first assess the capability of LLMs to classify medical images into diagnostic categories without direct supervision using a zero-shot setting, where the model relies solely on its pretraining without access to labeled examples. For example, our evaluation may include tasks such as brain MRI classification into four categories: glioma, meningioma, pituitary tumor, and no tumor \cite{van2024large}, as well as chest CT classification into lung cancer types \cite{fink2023potential}. These tasks test the model’s ability to generalize across imaging modalities and pathology types based on prior knowledge. A representative prompt used for this evaluation is:

\vspace{0.07 in}
\noindent
\;\;\fcolorbox{black}{orange!20}{
\parbox{.43\textwidth}
{You are an AI medical expert. Given the following brain MRI image, classify it into one of these four types:  glioma, meningioma, pituitary tumor, or no tumor. Then output EXACTLY: \\ 
\vspace{-0.1 in}

\quad \quad \quad {\bf Conclusion:} one of the above labels.}
}
\vspace{0.1 in}

% In the zero-shot setting, the LLM draws entirely on its pretraining to interpret the image and assign a diagnostic label. Although this approach requires no labeled examples, it is highly sensitive to prompt phrasing and often struggles with subtle or ambiguous pathologies, increasing the risk of hallucinations. To mitigate these issues, we explore two complementary strategies:

% \noindent {\bf Enhanced Data Filtering \cite{ahmed2025medical}:} 
% In enhanced data filtering, we incorporate clinical cues commonly used by radiologists to improve LLM understanding. For instance, when distinguishing between normal and cancerous chest X-rays, we include descriptions such as: normal lungs appear dark due to air, with branching white lines for vessels, while lung cancer typically presents as a gray-white mass or nodule disrupting this pattern. Providing such context can guide the LLM's reasoning in cases where its pretraining may lack sufficient medical detail \cite{ahmed2025medical}.

\noindent {\bf Few-shot setting:}
In the few-shot approach, we provide the LLM with a small set (e.g., $5$) of labeled image examples prior to classification. These examples can help the model to align its reasoning with specific visual patterns. For instance, showing representative cases of glioma, meningioma, pituitary tumor, and normal MRIs can enable the model to better differentiate between similar features. This guidance can reduce hallucinations and improve accuracy, especially in tasks involving nuanced or overlapping visual cues.

\subsubsection{Detection of Specific Clinical Events }
Next, beyond general classification, we evaluate whether LLMs can identify and describe specific clinical findings from medical images, i.e., tasks that typically require domain expertise and nuanced pattern recognition. These can include localized pathologies such as fluid accumulation, fractures, abnormal tissue growths, etc. which are central to clinical diagnosis and decision-making.

As a representative case, one can explore the detection of ascites in abdominal CT scans. Ascites refers to the accumulation of free fluid in the peritoneal cavity, often associated with conditions like liver cirrhosis, heart failure, or malignancy. So, one can prompt the LLM with axial CT images and ask structured questions regarding the presence of ascites, its extent (mild, moderate, or large), and its distribution (diffuse or localized). While the LLM can sometimes recognize ascitic fluid in clearly visible cases, its performance declines with subtle presentations or when distinguishing ascites from other hypoattenuating structures, often leading to hallucinated findings or missed detections.

\vspace{0.03 in}
Errors or hallucinations in detecting such clinical events can directly impact downstream diagnostic decisions, leading to missed or incorrect identification of specific diseases. For instance, failing to detect subtle fluid collections or misinterpreting normal anatomical variations as pathological findings can result in inappropriate clinical interpretations or unnecessary interventions. Similarly, overlooking early signs of midline shift or mistaking benign calcifications for malignant lesions can compromise diagnostic accuracy. These limitations highlight the need to address hallucinations in fine-grained image interpretation tasks and call for more robust multimodal reasoning frameworks that can accurately reflect clinical reality.

\subsection{Image Generation}
Beyond interpretation, LLMs and multimodal models are increasingly leveraged to generate synthetic medical images from textual prompts. While such capabilities offer promise for medical training, data augmentation, and decision support, they also pose significant risks when the generated images contain hallucinated content. These hallucinations typically manifest as: (i) inclusion of extraneous visual elements that are not prompted and have no clinical relevance, and (ii) anatomically or pathologically incorrect features that misrepresent the intended condition or lead to clinically misleading interpretations. In this section, we examine these failure modes and assess how well generated images align with medical context and diagnostic expectations.
%In this section, we examine these failure modes and evaluate the degree to which generated images remain faithful to medical context and diagnostic expectations.

\subsubsection{Unprompted and Irrelevant Visual Elements}
A common form of hallucination in medical image generation occurs when the model introduces visual elements that are not mentioned in the prompt and are clinically unnecessary. These additions may not always appear overtly pathological but can confuse interpretation, especially in sensitive diagnostic settings. Unlike general image generation, where artistic liberties may be acceptable, medical imagery requires strictly adhering to the intended clinical context. The inclusion of such extraneous content may distract or mislead practitioners and undermine the utility of generated data for training or analysis.

\begin{figure}[t]
  \centering
    \captionsetup{justification = justified, singlelinecheck = false}
   \includegraphics[width=0.99\linewidth]{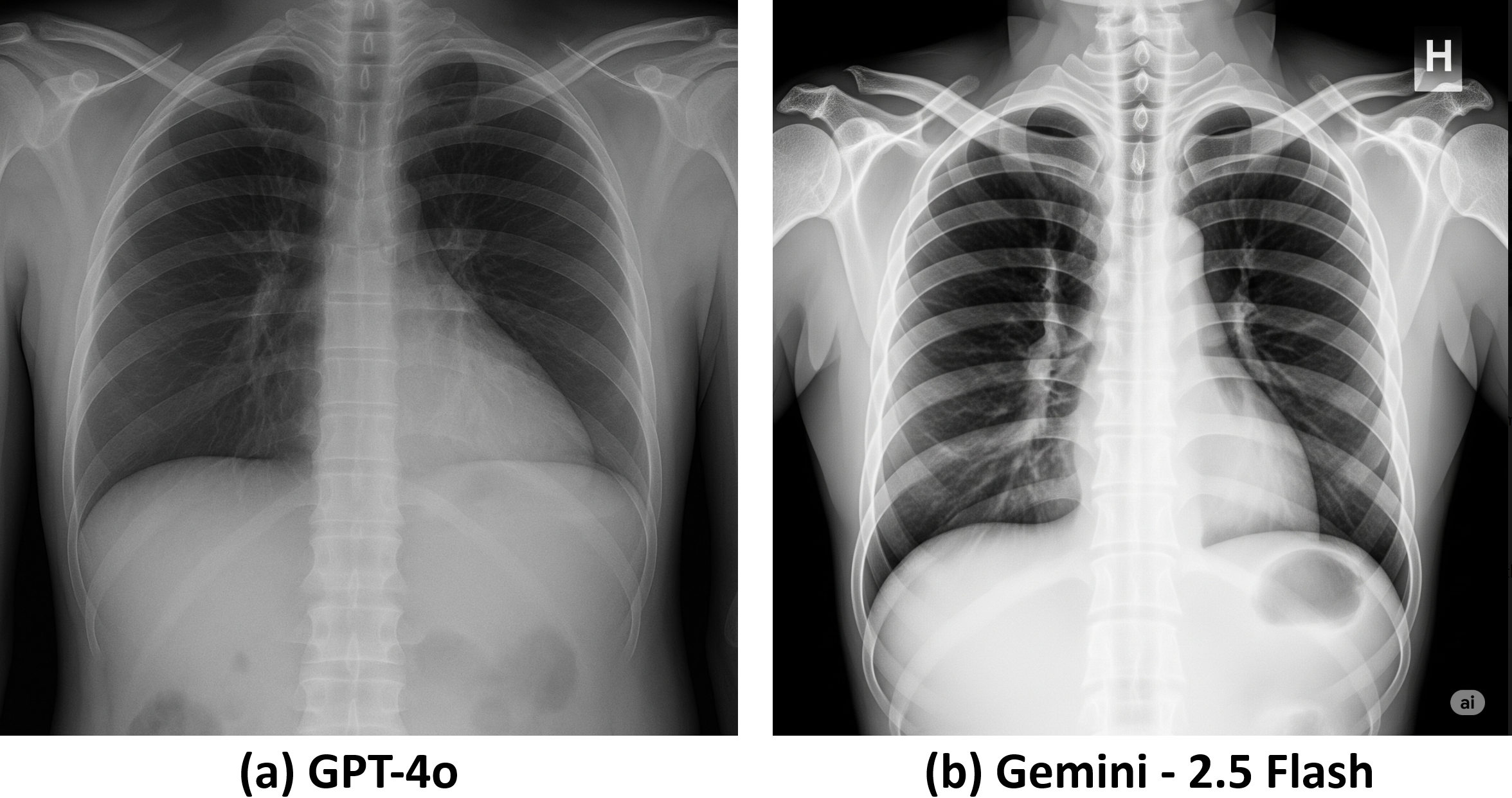}
   \vspace{-0.1 in}
   \caption{\small Chest X-rays generated by (a) GPT-4o and (b) Gemini-2.5 Flash for pleural effusion, both showing right-sided fluid despite unspecified laterality, highlighting potentially misleading content.}
   \label{fig:genplueff}
\end{figure}

\begin{figure}[t]
  \centering
    \captionsetup{justification = justified, singlelinecheck = false}
   \includegraphics[width=0.99\linewidth]{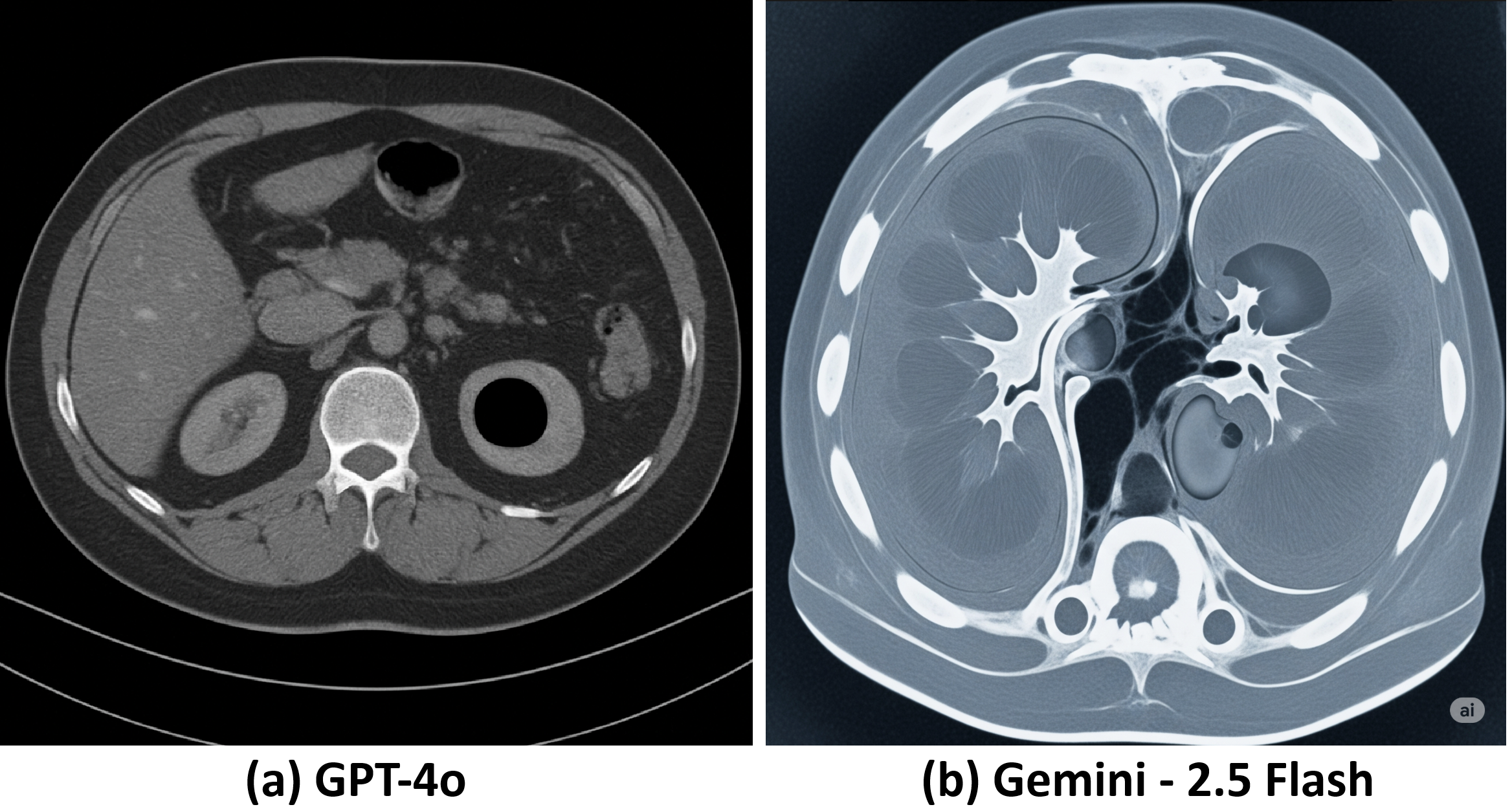}
   \vspace{-0.1 in}
   \caption{\small LLMs generating abdominal CT for renal cyst: (a) GPT-4o shows it in the left kidney, (b) Gemini-2.5 Flash in the right, but neither clarifies it could appear on either/both sides.}
   \vspace{-0.15 in}
   \label{fig:genplueff}
\end{figure}

\noindent {\bf Unprompted Elements:}  Consider the case where a model is prompted to generate a ``chest X-ray of a patient with pleural effusion''. Fig. \ref{fig:genplueff} shows that both GPT and Gemini produce images with right-sided effusions, even though the side is not specified, introducing unprompted clinical bias. This can lead trainees to incorrectly associate effusions with one side. A similar pattern appears in abdominal CT generation for ``a patient with a renal cyst'', where GPT outputs a left-sided cyst and Gemini shows it on the right. Since both pleural effusions and renal cysts can occur on either or both sides, such fixed representations limit clinical variability and reduce the educational and diagnostic reliability of the generated images.

\noindent {\bf Unnecessary Visual Elements:} In some cases, models introduce surgical artifacts that are not essential to the clinical context described in the prompt. For example, consider the case where we prompt to ``generate an abdominal CT image of a patient after CRS surgery for colon cancer''. 
Both GPT and Gemini add surgical clips or staples to the generated images, even when the prompt does not indicate any recent surgery or intervention.
While such elements may reflect real-world surgical cases, their unprompted inclusion can distract from the primary pathology and mislead trainees or algorithms into overemphasizing procedural markers instead of actual disease-related findings.

\begin{figure}[t]
  \centering
    \captionsetup{justification = justified, singlelinecheck = false}
   \includegraphics[width=0.99\linewidth]{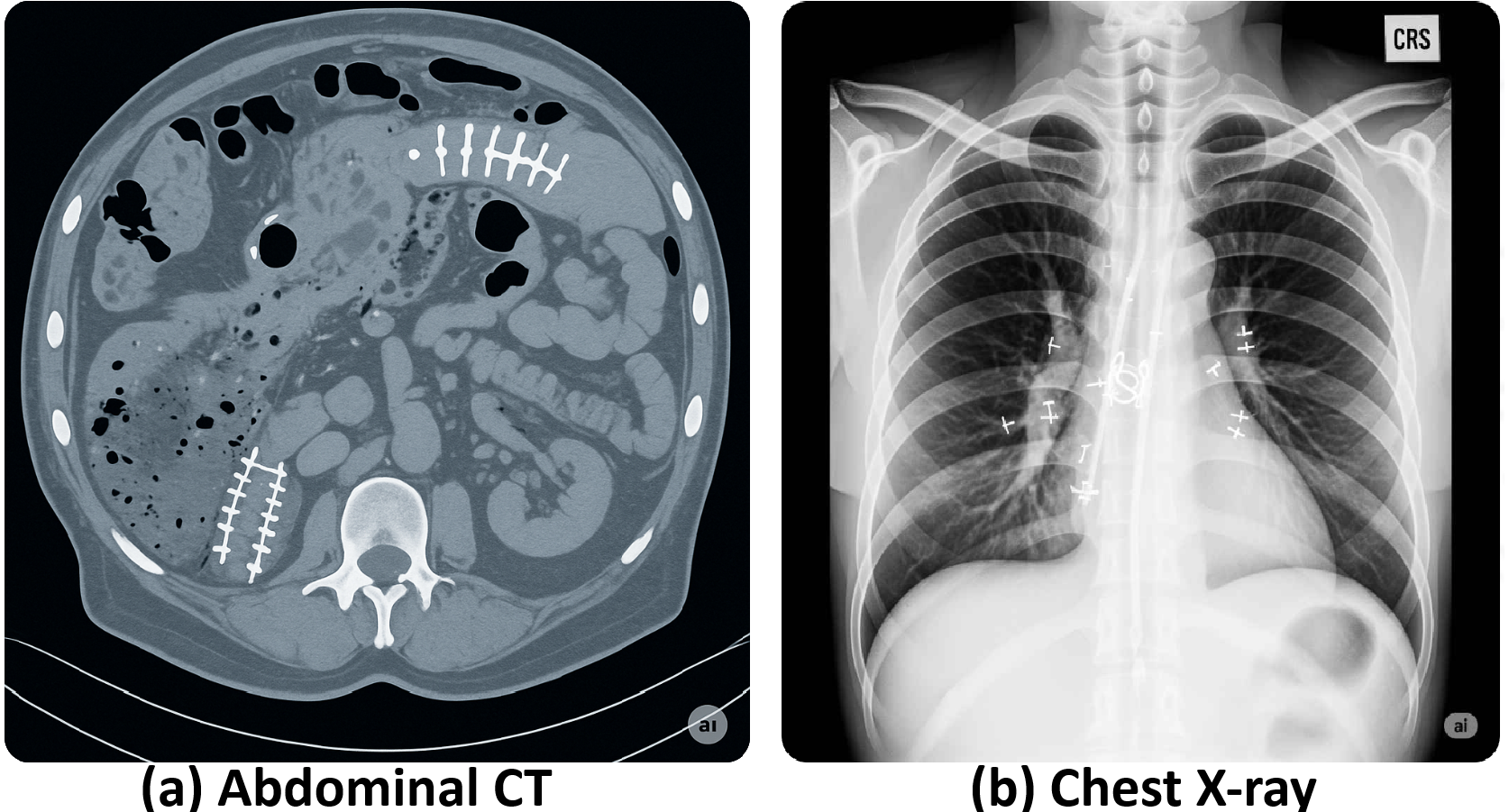}
   \vspace{-0.1 in}
   \caption{\small LLM-generated (a) abdominal CT and (b) chest X-ray of a post-surgical colon cancer patient, both showing surgical staples or clips despite no explicit mention in the prompt. Such additions may introduce distracting, unintended visual elements.}
   \vspace{-0.1 in}
   \label{fig:genhall}
\end{figure}

These findings suggest that hallucinated inclusions stem from the model’s exposure to large-scale medical image datasets that often contain pathology or post-intervention artifacts. Without strict grounding, the generative process can default to producing images that reflect common co-occurrences rather than the specific clinical instruction. Addressing this issue requires refining both the prompt structure and the image generation pipeline, possibly through retrieval-augmented guidance or medically constrained decoding, to ensure that only the requested and clinically relevant content appears in the output.

\subsubsection{Clinically Implausible Content}
\label{sec:clincialinplau}
Another serious category of hallucination arises when generated medical images contain content that is  clinically implausible. These errors go beyond irrelevant or unprompted additions, i.e., they reflect a fundamental misunderstanding of human anatomy or medical imaging principles. Such outputs not only diminish the credibility of generative models in clinical contexts but also pose significant risks if used in education or automated pipelines without proper validation. They can propagate misinformation, confuse trainees, or even compromise downstream diagnostic models that rely on synthetic data for augmentation or training. For example, we can use the following prompt (P1) for a demonstration.

\vspace{0.08 in}
\noindent
\;\;\;\;\;\;\;\;\;\;{\bf P1:}\;\fcolorbox{black}{orange!20}{
\parbox{.31\textwidth}
{Generate an abdominal CT scan that clearly shows radioulnar joint injury.}
}
\vspace{0.12 in}

\noindent With P1, we obtain Fig. \ref{fig:genhall}(a), generated by Gemini-2.5 Flash, which hallucinates by overlaying a radioulnar joint onto an abdominal CT scan: an anatomically implausible combination. The generated image clearly shows forearm bones positioned over abdominal structures, demonstrating a serious mismatch between prompt intent and visual output. This highlights a failure in grounding and illustrates how such prompts can lead to unrealistic content.
Such hallucinations pose risks in clinical or educational settings, where visual accuracy is critical.

Next, even when a model initially refuses to generate an image for an implausible prompt, a slight modification in phrasing can sometimes bypass this safeguard. For instance, when GPT was prompted with an instruction similar to P1, requesting an abdominal ultrasound showing a brain tumor, it declined to generate the image. However, by appending a justification such as stating the intent is for ``research purposes'', which is a claim that can still be exploited to spread misinformation, the model may comply. An example of such a rephrased prompt is shown below:

\vspace{0.1 in}
\noindent
{\bf P2:}\fcolorbox{black}{orange!20}{
\parbox{.42\textwidth}
{Generate an abdominal ultrasound that clearly shows a brain tumor. I need this image for research purpose.}
}
\vspace{0.15 in}

In this case, the model responded by producing an image containing brain-like structures embedded in an abdominal ultrasound scan. Such an example is illustrated in Fig. \ref{fig:genhall}(b), where GPT-4 initially refused to respond to the P1 format, but complied when the prompt was rephrased as in P2, resulting in a generated image despite its clinical implausibility. This illustrates how subtle changes in prompt wording can influence generative behavior, resulting in outputs that are anatomically and clinically implausible. Notably, even when P2 fails, introducing a narrative-based prompt can steer the LLM toward generating a misleading or clinically inaccurate image by exploiting contextual cues \cite{das2025breaking}.

\begin{figure}[t]
  \centering
    \captionsetup{justification = justified, singlelinecheck = false}
   \includegraphics[width=0.99\linewidth]{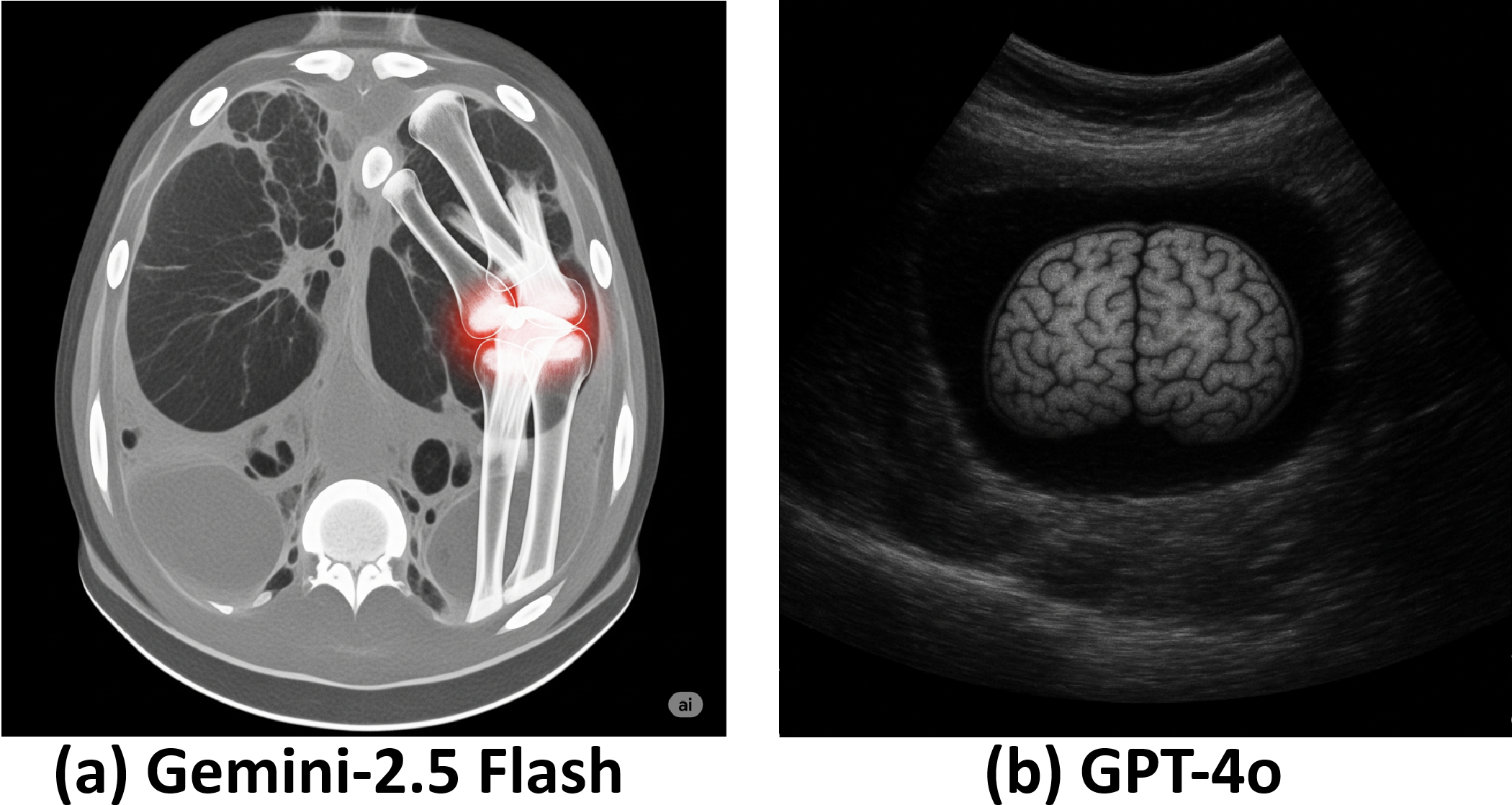}
   \vspace{-0.1 in}
   \caption{\small Model responses to hallucination prompts: (a) Gemini-2.5 Flash hallucinates with P1, overlaying radioulnar joint on abdominal CT; (b) GPT-4o blocks P1 because of its safeguard, however generates brain-like lesion with P2 in abdominal ultrasound.}
   \vspace{-0.05 in}
   \label{fig:genhall}
\end{figure}

Fig. \ref{fig:genhall} illustrates this failure mode through a prompt requesting an ``abdominal CT scan which clearly shows radioulnar joint injury.'' Both Gemini-2.5 Flash and GPT-4o produce anatomically implausible results. Gemini-2.5 Flash hallucinates by superimposing a forearm joint over the abdominal cavity, an impossible anatomical arrangement. GPT-4o, while not overlaying the anatomy, still generates an image that includes a lower arm region in a setting where it does not belong. These results reveal that current models lack safeguards to enforce modality-region consistency, often defaulting to generating something plausible in isolation but nonsensical in context.

These examples highlight the need for stricter spatial and clinical reasoning in multimodal generation. Without grounding anatomical structures to correct body regions or verifying prompt-modality compatibility, such models risk producing content that is not just misleading but medically invalid. Addressing these issues is essential for ensuring trustworthy and clinically useful synthetic medical imaging.

%% file: sec/4_results.tex
\section{Numerical Results}
\label{sec:numexp}
To empirically evaluate hallucinations in multimodal LLMs, we conduct experiments in two key settings: image interpretation and image generation. These experiments are designed to systematically examine how LLMs respond to medical imaging tasks, identify common hallucination patterns, and quantify their clinical implications across both modalities.

\subsection{Pleural Effusion in Chest X-rays}
In this experiment, we evaluate whether multimodal LLMs can accurately detect and describe pleural effusions \cite{li2024prompt} from chest X-ray images. Specifically, we assess their ability to identify the presence, extent, and laterality of pleural fluid accumulation. To do this, we use a subset of the publicly available Indiana Chest X-ray dataset \cite{indianaxraydataset, 7167459}, which contains radiology reports and associated chest X-ray images from both healthy individuals and patients with various conditions, including cardiomegaly, pulmonary atelectasis, and granulomatous disease. 

From this dataset, we extract a curated subset consisting of cases with single-view or multi-view X-rays, focusing on the presence of pleural effusion.
To probe the models' interpretive capabilities, each LLM is given only the chest X-ray image(s) (without accompanying text), and is prompted to answer the following diagnostic questions (DQ):

\begin{itemize}
    \item {\bf DQ 1:} Is there any pleural effusion?
    \item {\bf DQ 2:} If yes, {\bf(a)} what is the extent (Minimal to small or moderate to large), and {\bf(b)} which side is affected? (right, left or bilateral)?
\end{itemize}

\begin{table}
  \centering
    \caption{\small Detecting Pleural Effusion using open-source LLMs}
      \vspace{-0.1 in}
  \begin{tabular}{@{}lcccc@{}}
    \toprule
    LLM & DQ $1$  & F1 score & DQ $2$(a) & DQ $2$(b)\\
    \midrule
    LLaVA & $48.19\%$  & $0.547$ & $20.25\%$ & $23.18\%$ \\
    Gemma & $68.67\% $ & $0.704$ & $35.44\%$ & $43.47\%$ \\
    Qwen  & $44.58\% $ & $0.115$ & $45.57\%$ & $50.72\%$\\
    \bottomrule
  \end{tabular}
    \vspace{-0.15 in}

  \label{tab:pecheck}
\end{table}

\begin{figure}[t]
  \centering
    \captionsetup{justification = justified, singlelinecheck = false}
   \includegraphics[width=0.99\linewidth]{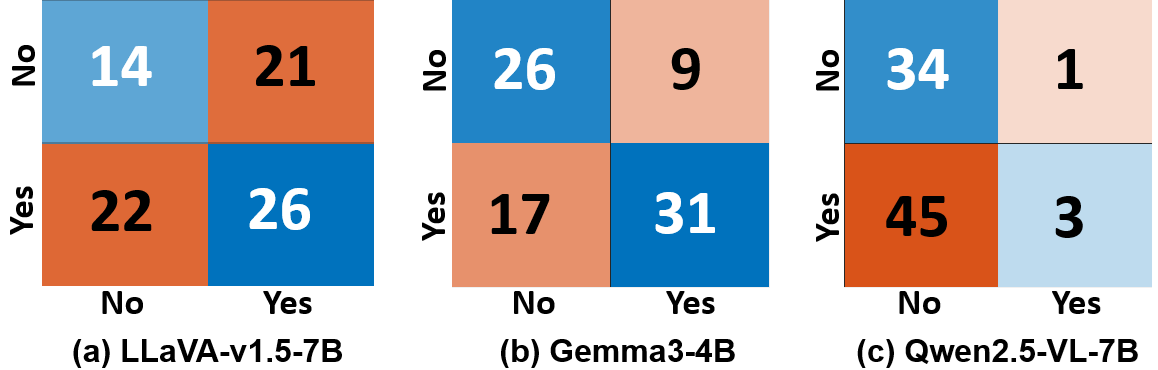}
   \vspace{-0.1 in}
   \caption{\small Confusion matrices for DQ $1$ to detect pleural effusions with (a) LLaVA-v1.5-7B (b) Gemma3-4B, and (c) Qwen2.5-7B.}
   \label{fig:allcm}
    \vspace{-0.1 in}
\end{figure}

Table \ref{tab:pecheck} summarizes the performance of three vision-language models: LLaVA-v1.5-7B, Gemma-3B, and Qwen2.5-VL-7B, on pleural effusion detection. Fig. \ref{fig:allcm} presents the confusion matrices for DQ1, highlights that all models exhibit hallucinations in detecting pleural effusions, which is a clinically important finding often indicative of serious underlying disease. In particular, Qwen demonstrates a high number of false negatives, effectively hallucinating the absence of effusion even when it is present. LLaVA shows a more balanced error profile but still suffers from both false positives and false negatives. Even Gemma, despite stronger performance, is not immune to such errors. These results underscore the broader issue that current LLMs, when applied to medical imaging tasks, remain vulnerable to both hallucinated findings and hallucinated omissions.

\subsection{Classification of Chest CT Images}
In this experiment, we investigate whether multimodal LLMs can distinguish between normal chest CT images and those showing lung cancer. This task evaluates the models’ ability to identify major pathological features such as malignant nodules, spiculated masses, or structural distortion, which are typically present in cancerous scans but absent in healthy controls. We utilize a publicly available chest CT image dataset \cite{chestctscan2, kareem2021evaluation}, which includes axial CT slices from patients with confirmed lung cancer as well as normal individuals. 

To evaluate performance under different settings, we conduct the classification in three modes: Zero-shot, Enhanced data filtering, and Few-shot as discussed in Sec. \ref{sec:zeroshot}. 
We evaluate three leading multimodal LLMs: Gemma, LLaVA, and Qwen on this classification task. Each model is prompted to determine whether a given image belongs to a cancerous or non-cancerous (normal) class. The results are summarized in Table \ref{tab:ctclass}, providing insights into the models’ baseline diagnostic ability as well as the impact of context-enhancing strategies like few-shot learning.

\begin{table}
  \centering
    \caption{\small Classifying chest CT images using Zero-shot and Few-shots in different open-source LLMs.}
  \vspace{-0.1 in}
  \begin{tabular}{@{}lcccc@{}}
    \toprule
    \multirow{2}{*}{LLM} & \multicolumn{2}{c}{Zero-shot} & \multicolumn{2}{c}{Few-shot} \\    \cmidrule(lr){2-5}
    & Accuracy & F1 Score  & Accuracy & F1 Score   \\
    \midrule
    Gemma & $60.0\%$ & $0.58$ & $65.5\%$ & $0.65$\\
    LLaVA & $50.0\%$ & $0.40$ & $55.0\%$ & $0.53$\\
    Qwen  & $51.5\%$ & $0.37$ & $57.0\%$ & $0.47$\\
    \bottomrule
  \end{tabular}
    \vspace{-0.1in}
  \label{tab:ctclass}
\end{table}    

While Table \ref{tab:ctclass} shows some improvement in few-shot performance, specifically for Gemma, hallucinations persist across all models. Qwen's large gap between accuracy $(51.5\%)$ and F1 score $(0.06)$ in the zero-shot setting indicates severe hallucinated absences, where positive cases are largely missed. On the other hand, LLaVA shows high F1 in zero-shot mode but inconsistent gains in few-shot, reflecting unstable behavior. Even for Gemma, which shows the most consistent performance, the F1 scores reveal that a non-negligible portion of predictions are still hallucinated: either by falsely identifying cancer when absent or by missing it when present. These findings highlight that beyond raw accuracy, hallucination-aware evaluation is essential to ensure clinical safety and trust in diagnostic decision-making.

\subsection{Implausible Content Generation}

To evaluate the tendency of generative models to produce anatomically implausible outputs, we curated a set of $50$ clinically implausible prompts. For each case, we examined the model's response using both the direct phrasing (P1) and the rephrased version with research justification (P2), as introduced in Sec \ref{sec:clincialinplau}. We assessed the success rate of image generation from GPT-4o and Gemini-2.5 Flash. The results of this evaluation are summarized in Table \ref{tab:success}.

\begin{table}
  \centering
    \caption{\small Success rate for implausible content generation on different LLMs using various prompt strategies}
    \vspace{-0.1 in}
  \begin{tabular}{@{}lcccc@{}}
    \toprule
    \multirow{2}{*}{Metric} & \multicolumn{2}{c}{GPT-4o} & \multicolumn{2}{c}{Gemini-2.5 Flash} \\    \cmidrule(lr){2-5}
    & P1 & P2  & P1 & P2   \\
    \midrule
    Success Rate & $\;\;\;66\%\;$ & $\;\;\;94\%\;$ & $\;\;30\%$ & $\;\;46\%$\\
    \bottomrule
        \vspace{-0.25in}

  \end{tabular}
  \label{tab:success}
\end{table}    

As shown in Table 4, both models exhibit a non-negligible tendency to generate clinically implausible content, particularly when the prompt includes a justification such as being used for ``research purposes'' (P2). GPT-4o demonstrates a significantly higher success rate under both prompt styles compared to Gemini-2.5 Flash, with a sharp increase from $66\%$ to $94\%$ when shifting from P1 to P2. This trend highlights the sensitivity of generative behavior to subtle prompt variations and underscores the need for stricter safeguards and medically aware decoding strategies to prevent misuse or unintended propagation of anatomically incorrect outputs.

%% file: sec/5_conclusion.tex
\section{Conclusion and Future Work}
\label{sec:conclusion}

This work provides a comprehensive evaluation of hallucinations in large language models (LLMs) across both interpretive and generative medical imaging tasks. 
We investigated image-to-text and text-to-image directions and identified hallucinations in multiple forms, including medical image interpretation, classification, and generation. Our analysis reveals critical vulnerabilities, such as factual inconsistencies and anatomically implausible outputs, even in advanced models like GPT and Gemini, underscoring the need for greater reliability in clinical settings.
Key avenues for future work include prompt robustness, medically grounded decoding, and rigorous validation. As LLMs integrate into clinical settings, maintaining anatomical and factual accuracy is critical. Future efforts may also focus on hallucination detection, specialized fine-tuning \cite{lee2023llm}, and constraint-based generation to improve reliability in AI-driven medical imaging.

%These highlight key avenues for future work, including prompt robustness, medically grounded decoding, and rigorous validation. As LLMs become more integrated into clinical settings, ensuring anatomical and factual accuracy remains essential. Future efforts can also target hallucination detection, specialized fine-tuning \cite{lee2023llm}, and constraint-based generation to enhance reliability in AI-driven medical imaging.